%% file: Hypernuclei_Recalibration_v2.2.tex
\documentclass[a4paper,10pt,twoside]{cpc-hepnp}

\usepackage{multicol}
\usepackage{graphicx}
\usepackage{subcaption}
\usepackage{booktabs}
\usepackage{amssymb,bm,mathrsfs,bbm,amscd}
\usepackage[tbtags]{amsmath}
\usepackage{lastpage}
\usepackage{CJKutf8}

\usepackage[colorlinks,linkcolor=blue,anchorcolor=blue,citecolor=blue,urlcolor=blue,filecolor=black]{hyperref}
\usepackage{cancel}
\usepackage{float}
\usepackage{lineno}
\usepackage{multirow}
\usepackage{longtable}

\begin{document}
%\begin{CJK*}{GBK}{song}
\begin{CJK}{UTF8}{gbsn}

%\fancyhead[c]{\small Chinese Physics C~~~Vol. 37, No. 1 (2019) 010201} \fancyfoot[C]{\small 010201-\thepage}

%\footnotetext[0]{Received 14 August 2019}

%Main text
\title{
Recalibration of the binding energy of hypernuclei measured in emulsion experiments and its implications
\thanks{
Supported by the Key Research Program of the Chinese Academy of Science (XDPB09), the National Natural Science Foundation of China (11890714, 11775288, 11421505, 11520101004), the China Scholarship Council (201704910615), the U.S. Department of Energy, Office of Science (DE-FG02-89ER40531, DE-SC-0012704)
}
}

\author{
Peng Liu$^{1, 2, 3}$
\quad Jinhui Chen$^{4;1)}$\email{chenjinhui@fudan.edu.cn}%
\quad Declan Keane$^{5}$
\quad Zhangbu Xu$^{3,6}$
\quad Yu-Gang Ma$^{4, 1}$
}
\maketitle

\address{
$^1$ Shanghai Institute of Applied Physics, Chinese Academy of Sciences, Shanghai 201800, China\\
$^2$ University of Chinese Academy of Sciences, Beijing 100049, China\\
$^3$ Brookhaven National Laboratory, Upton, New York 11973, USA\\
$^4$ Key Laboratory of Nuclear Physics and Ion-beam Application (MOE), Institute of Modern Physics, Fudan University, Shanghai 200433, China\\
$^5$ Kent State University, Kent, Ohio 44242, USA\\
$^6$ Shandong University, Qingdao, Shandong 266237, China\\
}

\begin{abstract}
The $\Lambda$ separation energy for $\Lambda$-hypernuclei, denoted $B_\Lambda$, measured in 1967, 1968, and 1973 are recalibrated using the current best mass estimates for particles and nuclei. The recalibrated $B_\Lambda$ are systematically larger (except in the case of $^6_\Lambda$He) than the original published values by about 100 keV. The effect of this level of recalibration is very important for light hypernuclei, especially for the hypertriton. The early $B_\Lambda$ values measured in 1967, 1968, and 1973 are widely used in theoretical research, and the new results provide better constraints on the conclusions from such studies.
\end{abstract}



%\footnotetext[0]{\hspace*{-3mm}\raisebox{0.3ex}{$\scriptstyle\copyright$}2013
%Chinese Physical Society and the Institute of High Energy Physics of the Chinese Academy of Sciences and the Institute of Modern Physics of the Chinese Academy of Sciences and IOP Publishing Ltd}%

\begin{multicols}{2}

\section{Introduction}
A hypernucleus contains one or more strange quarks, and in the most common type of hypernucleus, a neutron is replaced by a $\Lambda$ hyperon.  The first hypernucleus was discovered by Marion Danysz and Jerzy Pniewski in 1952 in a balloon-flown emulsion plate {\cite{Hypernuclei_1953}}. Many other hypernuclei were observed during the following few years in emulsion experiments{\cite{Hypernuclei_1958, Hypernuclei_1958_2}}. The hypertriton is the lightest hypernucleus, and is composed of a proton, a neutron, and a $\Lambda$ hyperon. The antimatter partner of the hypertriton was discovered in a heavy ion collision experiment by the STAR collaboration at the Relativistic Heavy Ion Collider (RHIC) located at Brookhaven National Laboratory in 2010{\cite{STAR2010}} and confirmed in 2015 by the ALICE collaboration at the Large Hadron Collider (LHC) located at CERN{\cite{ALICE2016}}. The hyperon-nucleon ($YN$) interaction plays an important role in understanding the strong nuclear force {\cite{Botta_2012, Muller_2000}}, and since hyperons may exist in the core of neutron stars{\cite{Hyperon_NS_2016}}, the $YN$ interaction is also of importance for the study of neutron star properties{\cite{strangeness_2016, HyperonPuzzle_2015, Hyperon_NS_2016}}. Hyperon-nucleon scattering would be a good tool to explore the $YN$ interaction, however it is very challenging to obtain stable hyperon beams due to the hyperon's very short lifetime. Hypernuclei are a natural $YN$ interaction system and thus their lifetime and binding energy have a direct connection to the strength of $YN$ interaction{\cite{Overbinding_2018, Beane_2013, STAR_2018}}. A precise determination of hypernuclear lifetime and binding energy can serve as critical inputs for theoretical study of the strong force and of neutron star interiors{\cite{Botta_2012, Muller_2000, Hyperon_NS_2016, strangeness_2016, HyperonPuzzle_2015, Overbinding_2018, Beane_2013, NS_phys_2004, NS_phys_2017, CHEN2018}}.

Although data on hyperon-nucleon scattering is lacking, measurements of the $\Lambda$ separation energy $B_\Lambda$ for $\Lambda$-hypernuclei have been available through nuclear emulsion experiments{\cite{Hypernuclei_1958, Hypernuclei_1958_2, strangeness_2016, BL_1966, NPB1_1967, NPB4_1968, NPB52_1973}}. The separation energy $B_\Lambda$ is defined as  ($M_\Lambda + M_{\rm core} - M_{\rm hypernucleus}) c^2$, where $M_{\rm{hypernucleus}}$, $M_{\Lambda}$ and $M_{\rm{core}}$ are the mass of hypernucleus, the $\Lambda$ hyperon, and the nuclear core of the hypernucleus, respectively, and $c$ is the speed of light. The $B_\Lambda$ measurements provided by emulsion experiments need to be revisited because the masses of particles and nuclei used in the original publications are different from contemporary best estimates from the PDG (Particle Data Group){\cite{PDG_2018}} and the AMDC (Atomic Mass Data Center, located at the Institute of Modern Physics, Chinese Academy of Sciences, Lanzhou){\cite{Nucleimass_2017_1, Nucleimass_2017_2}}. A case in point is a recent improved measurement of the hypertriton's $B_{\Lambda}$ reported by the STAR collaboration in 2019{\cite{STAR_2019, STAR_2019_2}}, where the best estimate is significantly larger than the previous commonly-cited value published in 1973{\cite{NPB52_1973}}. However, the early measurements of $B_\Lambda$ have been used as critical inputs for theoretical research. For example, Ref.{\cite{Overbinding_2018}} applied the \cancel{$\pi$}EFT approach at LO to $s$-shell $\Lambda$ hypernuclei with precise few-body stochastic variational method calculations to address the over-binding problem of $^{5}_{\Lambda}$He.  In another relevant example, Ref.{\cite{HyperonPuzzle_2015}} considered two different models of three-body force which is constrained by the $\Lambda$ separation energy of medium-mass hypernuclei, but the authors obtained dramatically different results for the maximum mass of neutron stars using different models of the three-body force. Stronger constraints on the $YN$ interaction are necessary to properly understand the role of hyperons in neutron stars. From this point of view, it is timely and highly desirable to recalibrated these early measurements using the contemporary best mass estimates for particles and nuclei{\cite{PDG_2018, Nucleimass_2017_1, Nucleimass_2017_2}} to provide more accurate constraints for contemporary studies{\cite{strangeness_2016, Overbinding_2018, AFDMC_2014, AFDMC_2017}}.

\section{Techniques of recalibration}
In the emulsion experiments, $B_\Lambda$ was defined by {\cite{Hypernuclei_1958, Hypernuclei_1958_2}}:
\begin{equation}
\label{B_eq}
B_\Lambda = Q_0 - Q
\end{equation}

\begin{equation}
\label{Q0}
Q_0 = M_{F^{'}} + M_{\Lambda} - \sum_{i}{M_{i}}
\end{equation}
where $Q$ is the total kinetic energy released when a hypernucleus decays through a mesonic decay channel, $F^{'}$ represents the nuclear core of the hypernucleus, $M$ is the mass of a particle or nucleus, and the subscript $i$ refers to the $i^{\rm th}$ decay daughter. $Q$ was determined using the range-energy relation in emulsion{\cite{Hypernuclei_1958, Hypernuclei_1958_2, NPB1_1967, NPB4_1968, NPB52_1973, RE_1970}}, while $Q_{0}$ was directly determined from the masses of particles and nuclei available at the time of publication. A series of papers for nuclide masses was published at that time (in 1954{\cite{Nucleimass1954}}, 1960{\cite{Nucleimass1960_1,Nucleimass1960_2}}, 1962{\cite{Nucleimass1962}}, 1965{\cite{Nucleimass1965}}, 1971{\cite{Nucleimass1971}}, and 1977{\cite{Nucleimass1977}}), and the nuclide masses published in 1960{\cite{Nucleimass1960_1,Nucleimass1960_2}} were used in Ref.{\cite{BL_1966}} on $B_{\Lambda}$ measurements. Ref.{\cite{BL_1966}} shared the same corresponding author as Refs.{\cite{NPB1_1967, NPB4_1968, NPB52_1973}}, and therefore we assume here that the masses used in 1967{\cite{NPB1_1967}} and 1968{\cite{NPB4_1968}} were taken from the nuclide mass paper published in 1965{\cite{Nucleimass1965}}. We also assume here that the masses used in 1973{\cite{NPB52_1973}} were taken from the nuclide mass paper published in 1971{\cite{Nucleimass1971}}. The masses for $\pi^{-}$, proton, $\Lambda$, and the relevant nuclei used in the past and in 2019 are listed in Table \ref{table1}. From Table \ref{table1}, it is evident that the early masses of particles and nuclei are different from current values. Consequently, the $Q_{0}$ values used in 1967, 1968, and 1973 were such that the original published $B_\Lambda$ measurements were not as accurate as they could be. Fortunately, the early $B_\Lambda$ values can be recalibrated by comparing the difference in $Q_{0}$ between the old publications and modern numbers.  According to the masses listed in Table \ref{table1}, the $Q_{0}$ for light hypernuclei with mass number $A =$ 3 - 15 are calculated for specific decay channels in the years 1967, 1968, 1973, and 2019. Table \ref{table2} presents $Q_{0}$ and $\Delta Q_{0}$, where the latter is the difference between $Q_{0}$ in 2019 and $Q_{0}$ in the specific earlier year. $\Delta Q_{0}$ is used for recalibrating the $B_{\Lambda}$ measured in 1967, 1968, and 1973. The original $B_{\Lambda}$ is recalibrated for each decay channel listed in Table \ref{table2}. After recalibration, this paper provides more precise estimations of the $B_\Lambda$ values. Table \ref{table3} lists the original and recalibrated $B_\Lambda$ values for a combination of all available decay channels for the listed $\Lambda$-hypernuclei with mass numbers $A =$ 3-15.

We note that the early emulsion measurements in 1968 and 1973 benefited from a compensating effect in normalizing the $B_{\Lambda}$ values via measuring the mass of the $\Lambda$ hyperon with the decay daughter $\pi^{-}$ range of 1-2 cm in the same emulsion stack{\cite{NPB4_1968, NPB52_1973}}. Because the mass of $\Lambda$ hyperon and the $Q$ appear with opposite sign in equation \ref{B_eq}, it was argued that systematic errors arising from uncertainties in both the range-energy relation and the emulsion density had been largely offset by the normalization procedure and a small systematic error of 0.04 MeV was determined for the early measurements{\cite{NPB4_1968, NPB52_1973, Davis_1986, Davis_1992, Davis_2005}}. Although an identical $\Lambda$ mass was measured in 1968 and 1973, the differences of $B_\Lambda$ between measurements in these two years are large (up to 0.50 $\pm$ 0.17 (stat.) MeV{\cite{NPB52_1973}}), which is significantly larger than the systematic error of 0.04 MeV. The ranges of $\pi^{-}$ from $\Lambda$ decays in the emulsion experiments were chose to be 1-2 cm, a range interval that covers the great majority of $\pi^{-}$ from hypernuclear decays. Nevertheless, the distribution of $\pi^{-}$ ranges from $\Lambda$ decay and hypernuclear decay are different{\cite{NPB4_1968, NPB52_1973, RE_1970}}. This difference in $\pi^{-}$ range can also yield a difference in the measured $Q$ value as large as 0.43 $\pm$ 0.13 (stat.) MeV {\cite{RE_1970}}, and cannot ensure the deviations of measured $Q$ value for $\Lambda$ decay and hypernuclear decay are in the same direction{\cite{RE_1970}}. Recent precise measurements show that for p-shell hypernuclei, there is a discrepancy in the range of 0.4-0.8 MeV between early emulsion measurements and their measurements{\cite{Achenbach2017}}. The authors of Ref.{\cite{Achenbach2017}} also argued that the emulsion data significantly underestimated the systematic error, which is dependent on the specific hypernucleus. Based on the above statement, we believe that the compensating effect described above may not fully account for the systematic error, and a recalibration of the $Q_{0}$ differences seems to be a more reliable method. 

\end{multicols}

\input{table1.tex}
\input{table2.tex}
\input{table3.tex}

\begin{multicols}{2}

\section{Results and discussions}
We note that it is tempting to average the recalibrated early measurements for each hypernucleus to obtain a more precise best value. However, as we explained above, without a better understanding of the systematic uncertainty associated with emulsion measurements, it is not appropriate to perform a weighted average. From Table \ref{table3}, it is evident that the recalibrated $B_\Lambda$ values are systematically larger than the original estimates except in the case of $^{6}_{\Lambda}$He. Comparing with the original $^{3}_{\Lambda}$H $B_\Lambda$ = $\rm 0.13 \pm 0.05 (stat.)$ MeV which was published in 1973 and widely used in modern theoretical studies, the recalibrated $B_\Lambda$ = $\rm 0.27 \pm 0.08 (stat.)$ MeV is closer to the latest result, namely $B_\Lambda$ = $\rm 0.41 \pm 0.12(stat.) \pm 0.11(syst.)$ MeV published by the STAR collaboration in 2019{\cite{STAR_2019, STAR_2019_2}}. The latest precise measurement of $^{4}_{\Lambda}$H by A1 collaboration in 2016 is 2.157 $\pm$ 0.005 (stat.) $\pm$ 0.077 (syst.) MeV{\cite{A1_Collaboration_2016}}, which is also closer to our recalibrated values compared with the original values presented in 1973. In addition, in contrast to the original emulsion measurements, our recalibrated values for $^{7}_{\Lambda}$He and $^{7}_{\Lambda}$Li are also closer to the latest measurement $B_\Lambda (^{7}_{\Lambda}$He) = $\rm 5.55 \pm 0.10(stat.) \pm 0.11(syst.)$ MeV by the HKS collaboration{\cite{HKS_He7}} in 2016 and $B_\Lambda (^{7}_{\Lambda}$Li) = $\rm 5.85 \pm 0.13(stat.) \pm 0.11(syst.)$MeV{\cite{CSB_2017, FINUDA_Li7_Be9}} by the FINUDA collaboration in 2009 than original values of emulsion experiments. These numbers corroborate the expectation that all the recalibrated $B_\Lambda$ values presented in this paper are indeed better estimates than the early measurements in the light hypernuclei region. In recent years, collaborations at Jefferson Lab and the DA$\Phi$NE-FINUDA collaboration measured $B_{\Lambda}$ with good accuracy for heavier hypernuclei, namely $B_\Lambda (^{9}_{\Lambda}$Li) = $\rm 8.36 \pm 0.08(stat.) \pm 0.08(syst.)$ MeV{\cite{HKS_Li9}}, $B_\Lambda (^{9}_{\Lambda}$Be) = $\rm 6.30 \pm 0.10(stat.) \pm 0.10(syst.)$ MeV{\cite{CSB_2017, FINUDA_Li7_Be9}}, $B_\Lambda (^{10}_{\Lambda}$Be) = $\rm 8.60 \pm 0.07(stat.) \pm 0.16(syst.)$ MeV{\cite{HKS_Be10}}, $B_\Lambda (^{11}_{\Lambda}$B) = $\rm 10.28 \pm 0.2(stat.) \pm 0.4(syst.)$ MeV{\cite{CSB_2017}}, $B_\Lambda (^{12}_{\Lambda}$B) = $\rm 11.524 \pm 0.019(stat.) \pm 0.013(syst.)$ MeV{\cite{HKS_B12}}, $B_\Lambda (^{13}_{\Lambda}$C) = $\rm 11.0 \pm 0.4$ MeV{\cite{FINUDA_C13}}, and $B_\Lambda (^{15}_{\Lambda}$N) = $\rm 13.8 \pm 0.7(stat.) \pm 1.0(syst.)$ MeV{\cite{CSB_2017}}. Comparing these measurements with early emulsion measurements, some recent results indicate larger $B_{\Lambda}$. However, some of them show smaller $B_{\Lambda}$.

\begin{figure}[H]
\centering
	\begin{subfigure}{0.40\linewidth}
		\centering
		\includegraphics[width=1.0\textwidth]{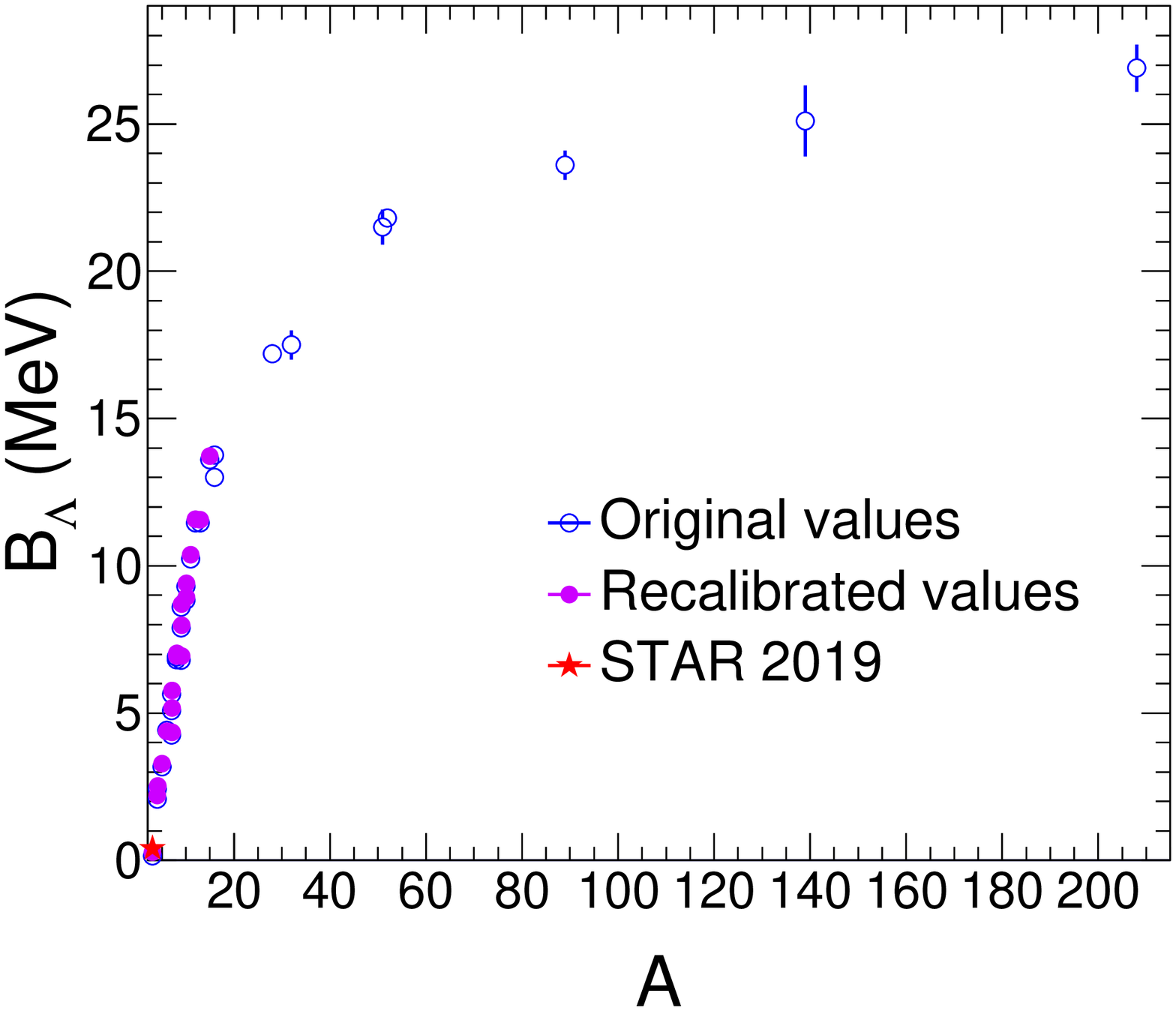}
		\label{Fig1a}
	\end{subfigure}
	\begin{subfigure}{0.40\linewidth}
		\centering
		\includegraphics[width=1.0\textwidth]{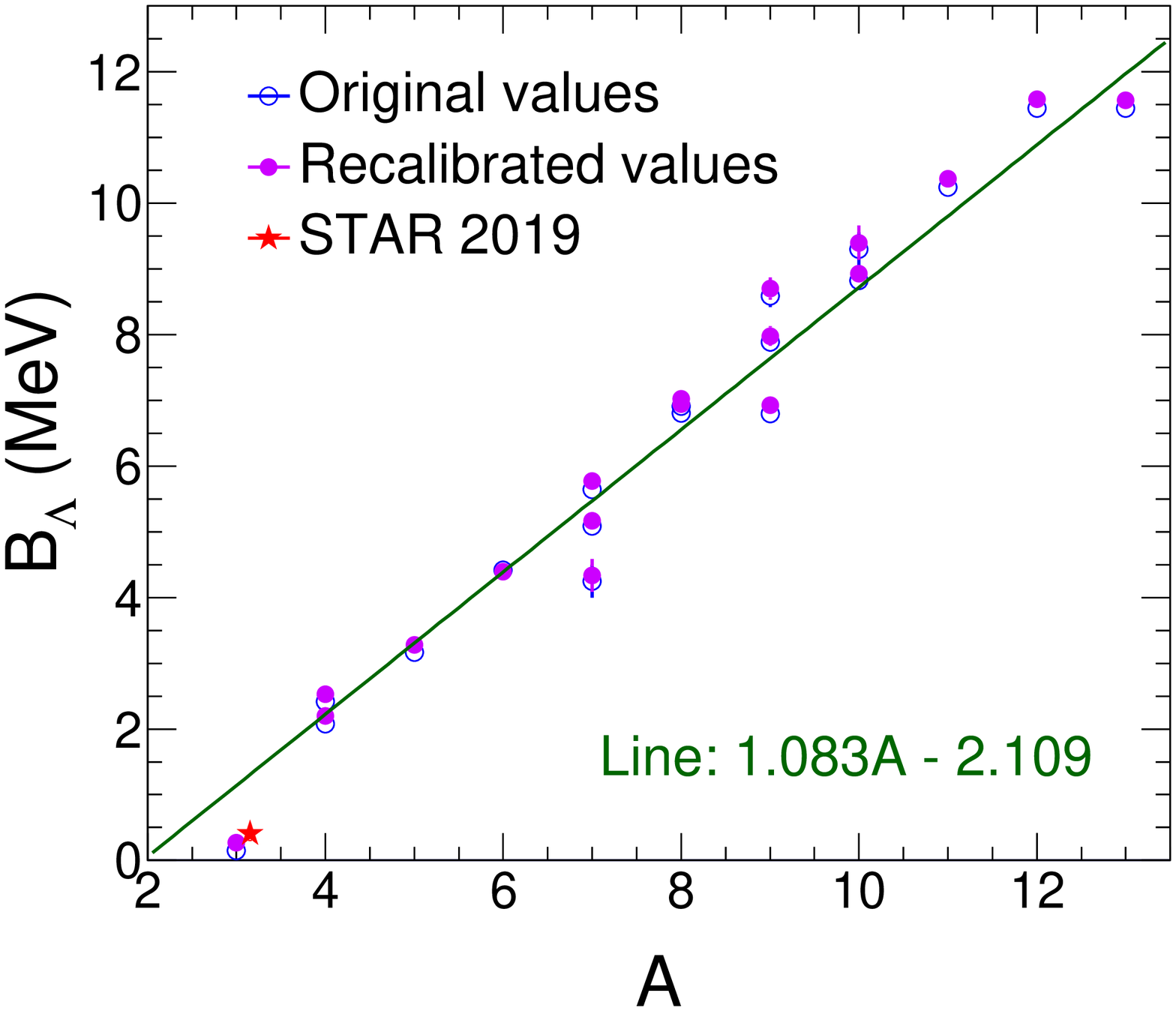}
		\label{Fig1b}
	\end{subfigure}
	\caption{The $\Lambda$ separation energy $B_{\Lambda}$ of hypernuclei as a function of mass number $A$. The original values and the recalibrated values are shown together with the latest measurement for the hypertriton by the STAR collaboration{\cite{STAR_2019}}. The error bars are the reported uncertainties. The caps and error bar shown for the STAR measurement are the systematic and statistical uncertainty, respectively. The right panel shows a magnified view. The STAR point is moved away from A=3 a bit to make it visible. A straight line is fitted to the recalibrated values in the range $A =$ 3 -15. The green line and the green text shown in this figure are the fit results.}
	\label{Fig1}
\end{figure}

\begin{figure}[H]
\centering
	\begin{subfigure}{0.40\linewidth}
		\centering
		\includegraphics[width=1.0\textwidth]{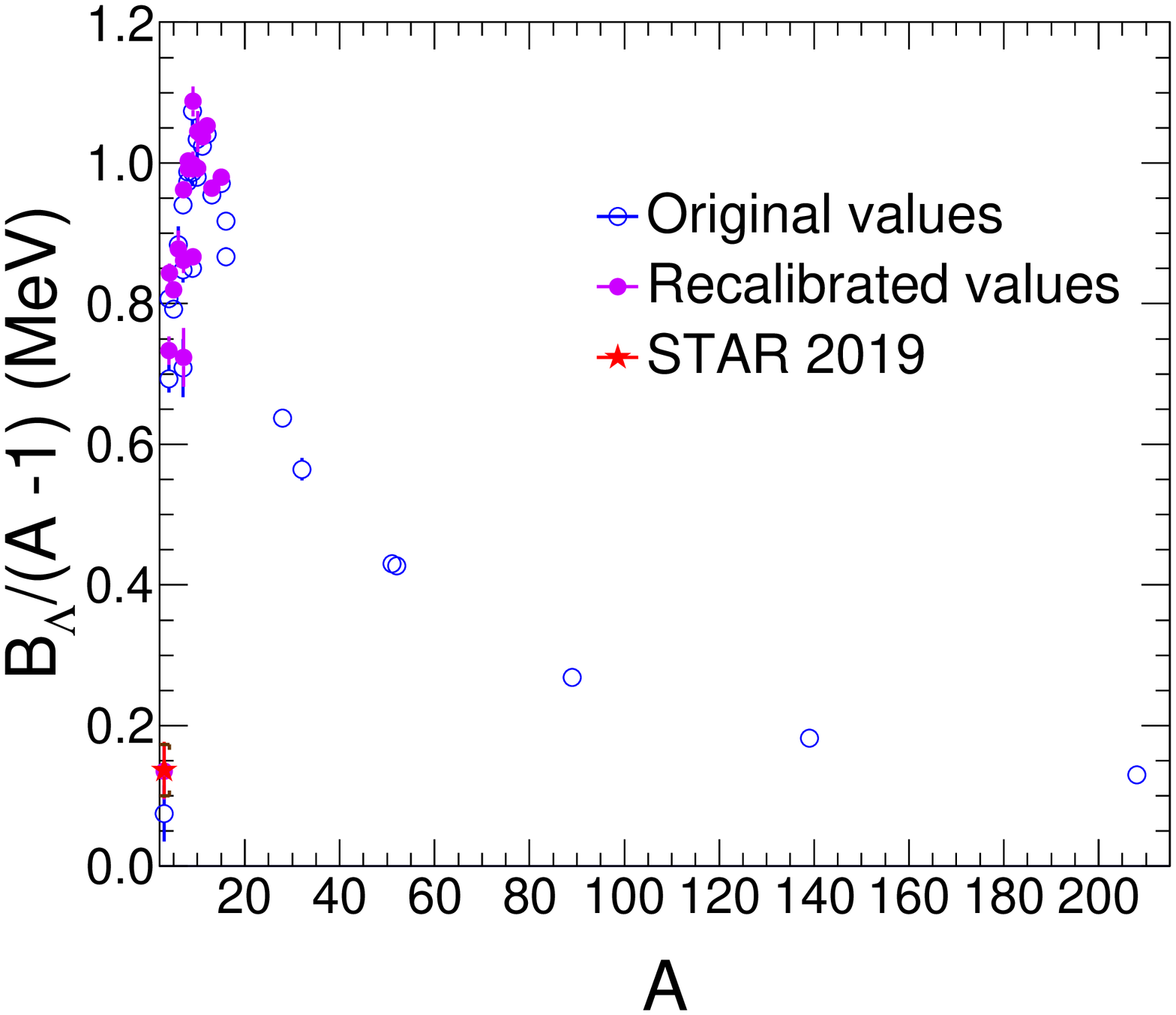}
		\label{Fig2a}
	\end{subfigure}
	\begin{subfigure}{0.40\linewidth}
		\centering
		\includegraphics[width=1.0\textwidth]{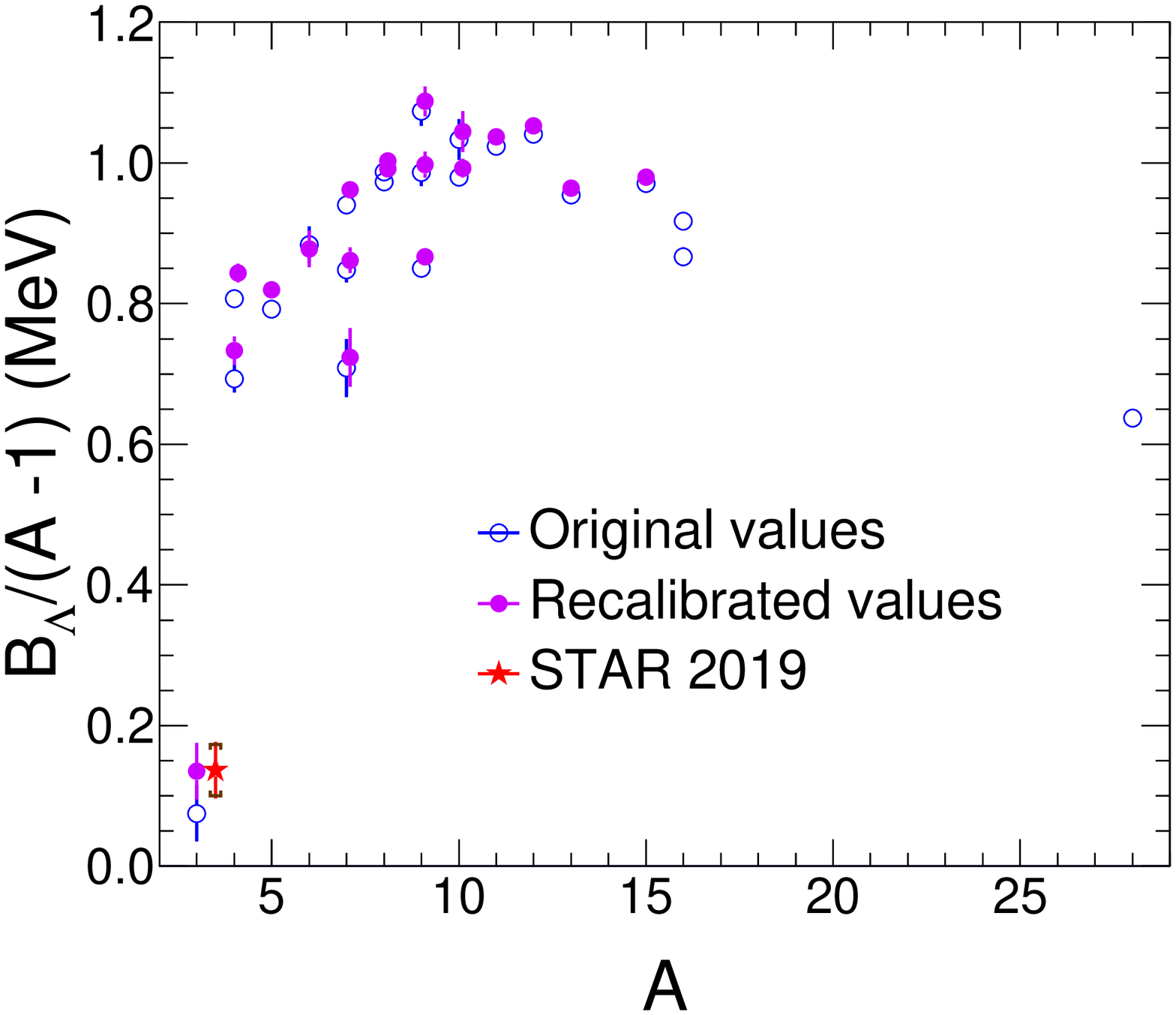}
		\label{Fig2b}
	\end{subfigure}
	\caption{The hypernuclear $\Lambda$ separation energy $B_{\Lambda}$ per baryon in the core of hypernuclei as a function of mass number $A$. The original values and the recalibrated values are shown together with the latest measurement for the hypertriton by the STAR collaboration{\cite{STAR_2019}}. The error bars are the reported uncertainties. The caps and error bar shown for the STAR measurement are the systematic and statistical uncertainty, respectively. The right panel shows a magnified view. The STAR point is displaced slightly from $A=3$ for visibility.}
	\label{Fig2}
\end{figure}

\begin{figure}[H]
\centering
	\begin{subfigure}{0.40\linewidth}
		\centering
		\includegraphics[width=1.0\textwidth]{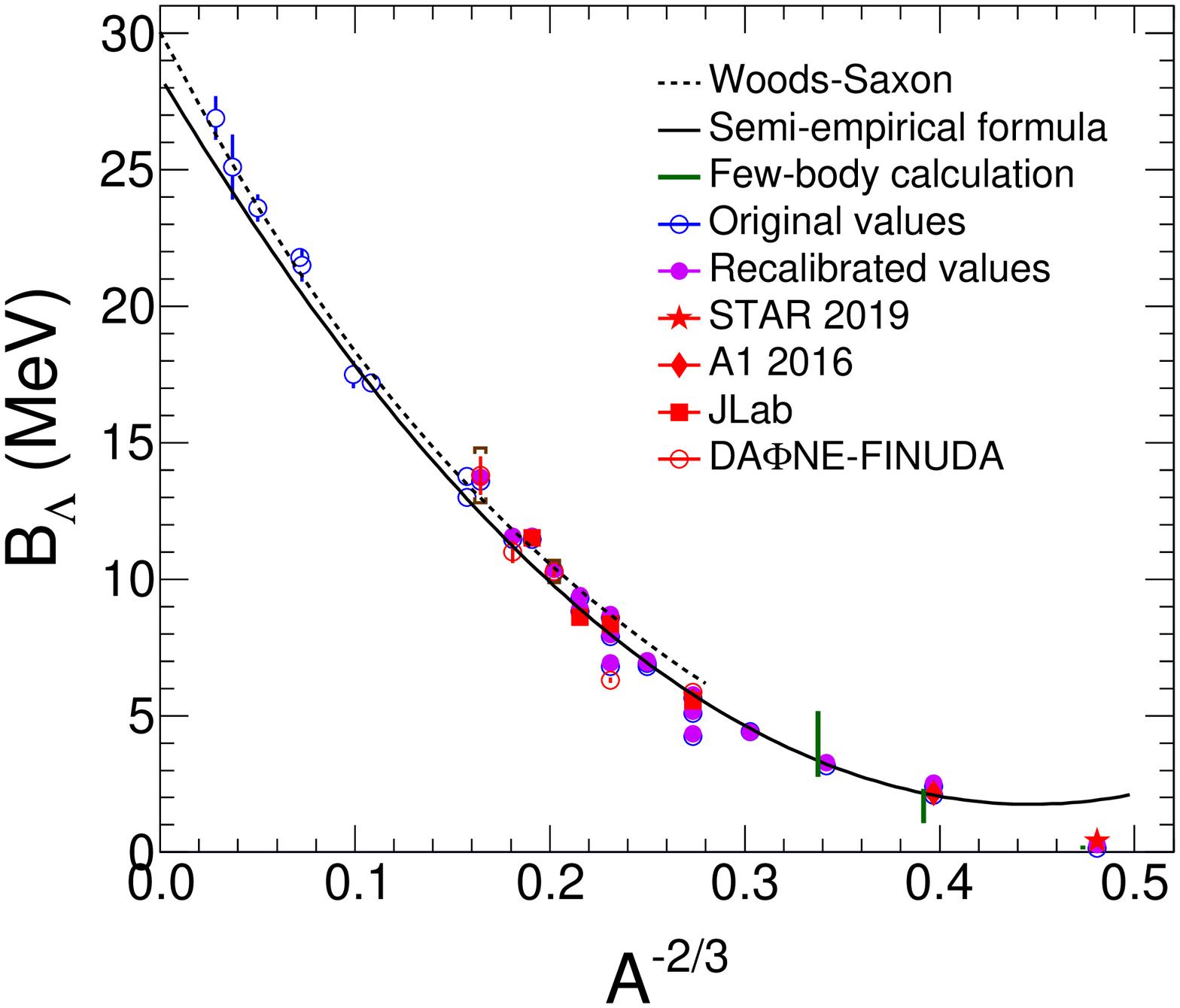}
		\label{Fig3a}
	\end{subfigure}
	\begin{subfigure}{0.40\linewidth}
		\centering
		\includegraphics[width=1.0\textwidth]{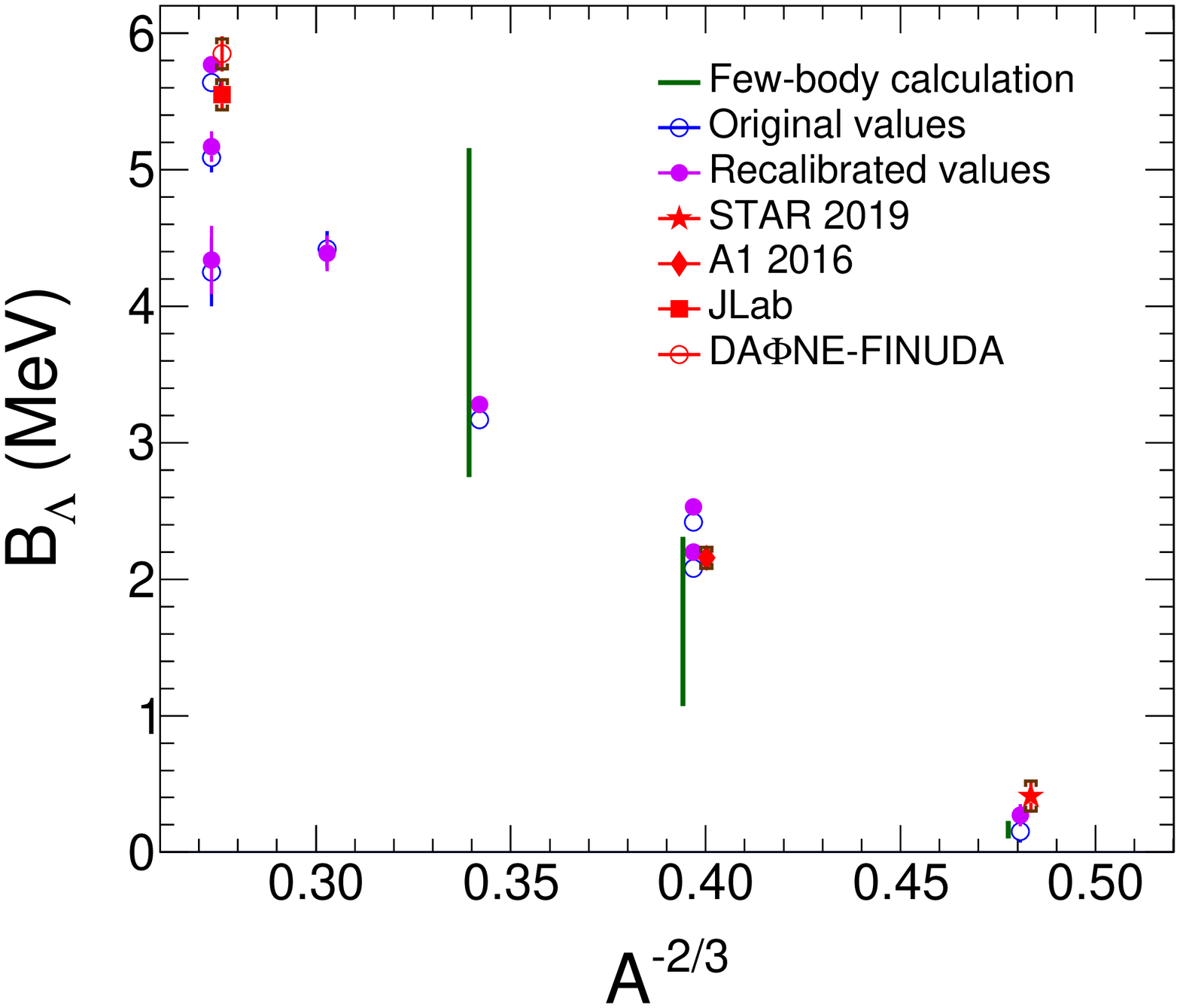}
		\label{Fig3b}
	\end{subfigure}
	\caption{The hypernuclear $\Lambda$ separation energy $B_{\Lambda}$ as a function of $A^{-2/3}$. The original values and the recalibrated values are shown together with the latest measurement for the $^{3}_{\Lambda}$H by the STAR collaboration{\cite{STAR_2019}}, measurement for the $^{4}_{\Lambda}$H by A1 collaboration{\cite{A1_Collaboration_2016}}, measurements for the $^{7}_{\Lambda}$He, $^{9}_{\Lambda}$Li, $^{10}_{\Lambda}$Be, and $^{12}_{\Lambda}$B by JLab{\cite{HKS_He7, HKS_Li9, HKS_Be10, HKS_B12}}, and measurements for the $^{7}_{\Lambda}$Li, $^{9}_{\Lambda}$Be, $^{11}_{\Lambda}$B, $^{13}_{\Lambda}$C, and $^{15}_{\Lambda}$N by DA$\Phi$NE-FINUDA collaboration{\cite{CSB_2017, FINUDA_Li7_Be9, FINUDA_C13}}. The error bars are the reported uncertainties. The caps and error bars shown for the STAR, A1, JLab, and DA$\Phi$NE-FINUDA measurements are the systematic and statistical uncertainty, respectively. The dashed black curve in the left panel was obtained by solving the Schr{\"o}dinger equation with a standard Woods-Saxon potential{\cite{strangeness_2016}} and the solid black curve is a semi-empirical formula{\cite{semi-empirical_2018}}. The green vertical lines nearby experiment points are several representative few-body calculations{\cite{Overbinding_2018, AFDMC_2017}}. The right panel shows a magnified view. In this panel, The markers for STAR, A1, JLab, and DA$\Phi$NE-FINUDA are displaced slightly from their corresponding mass numbers for visiblity.}
	\label{Fig3}
\end{figure}

The recalibrated $\Lambda$ separation energy $B_\Lambda$ of hypernuclei in 1973 (except for $^{7}_{\Lambda}$He, whose value dates from 1968) along with values{\cite{strangeness_2016}} for hypernuclei with mass number $A >15$ are collected in Figs. \ref{Fig1}, \ref{Fig2}, and \ref{Fig3}. Figure \ref{Fig1} shows $B_\Lambda$ as a function of hypernuclear mass number $A$. From Fig. \ref{Fig1}, it is evident that $B_{\Lambda}$ dramatically increases with mass number up to about $A \sim 15$. As $A$ becomes larger, $B_{\Lambda}$ increases more slowly and indicates a trend towards saturation in the limit of very large $A$. As shown in the right panel of Fig. \ref{Fig1}, a straight line provides a good fit to the recalibrated $B_{\Lambda}$ in the region of light hypernuclei, i.e., $A < 15$. 

Figure \ref{Fig2} investigates the strength of the interaction between a nucleon in the core of a hypernucleus and the bound $\Lambda$. From Fig. \ref{Fig2}, it is evident that the strength of the interaction between a nucleon and the $\Lambda$ dramatically increases with $A$ and then decreases. At very large $A$, it shows a tendency to flatten out. From the right panel of Fig. \ref{Fig2}, it is evident that $B_{\Lambda}/(A- 1)$ reaches a maximum between $A = 8$ and $A = 12$.  Fig. \ref{Fig3} presents $B_\Lambda$ versus $A^{-2/3}$. The dashed black curve is the solution of the Schr{\"o}dinger equation with the standard Woods-Saxon potential{\cite{strangeness_2016}}, which describes the medium and heavy mass range. A semi-empirical formula based on Fermi gas model is also employed to find the $B_{\Lambda}$ of light, medium, and heavy hypernuclei{\cite{semi-empirical_2018}} as shown in Fig. \ref{Fig3}. Although this semi-empirical formula has a good performance at mid-mass range, it is not good at fitting the experiment data in light and heavy mass range. On the other hand, both Woods-Saxon potential and semi-empirical formula does not take into account the CSB effect. The right panel of Fig. \ref{Fig3} zooms-in on the region $A\leq 7$, where theoretical calculations span a wide range as shown in the right panel of Fig. \ref{Fig3} and consequently our recalibration becomes more important.

The recalibrated values are systematically larger (except in the case of $^6_\Lambda$He) than original values by about 100 keV which is mainly from the contribution of $\Lambda$ hyperon mass difference between the modern values and the early emulsion values. This effect of about 100 keV is more significant for light hypernuclei, especially for the $^{3}_{\Lambda}$H, since its $B_\Lambda$ is very small compared with heavy hypernuclei. These larger $B_{\Lambda}$ of light hypernuclei obtained through recalibrating the emulsion data and recent experimental measurements will help us understand the puzzle of shortened lifetime of $^{3}_{\Lambda}$H{\cite{shorter_lifetime}}. The latest compilation of measurements yields a $^3_\Lambda$H lifetime shorter than the free $\Lambda$ lifetime {\cite{STAR_2018,ALICE2019}}. A calculation in which the closure approximation was introduced to evaluate the wave functions by solving the three-body Faddeev equations, indicates that the $^3_\Lambda$H lifetime is $(19 \pm 2)\%$ smaller than that of the $\Lambda${\cite{lifetime_2019}}. The shorter lifetime is consistent with a larger $\Lambda$ separation energy in $^{3}_{\Lambda}$H. The significant change in $B_\Lambda$ of $^3_\Lambda$H will also improve the understanding of other hypernuclei{\cite{over_binding}}. We also learn from Table \ref{table3} that $B_\Lambda$ for hypernuclei with the same mass number $A$ but different electric charge are significantly different, i.e., the Charge Symmetry Breaking{\cite{CSB_2017}} (CSB) effect. Theoretical studies are particularly needed to address the CSB effect.

\end{multicols}

\input{table4}

\begin{multicols}{2}

We also investigate the difference between $B_{\Lambda}$ of hypernuclei and the corresponding binding energy of the last neutron and proton ($S_{n}$ and $S_{p}$) of ordinary nuclei, as shown in Table \ref{table4}. The $B_{\Lambda}$ 
increases with $A$, but $S_{n}$ and $S_{p}$ feature a significantly different behavior. This difference means that a $\Lambda$ hyperon plays a role that is different from that of a nucleon in a nucleus. On the other hand, this difference may be related to the rich structure of the nuclear core. For example, the Gaussian expansion method provides an accurate structure calculation of light hypernuclei by treating them as three and/or four clusters{\cite{structure_2009}}.

\section{Summary}
In summary, early measurements of $\Lambda$ separation energy $B_\Lambda$ for $\Lambda$ hypernuclei published in 1967, 1968, and 1973 are recalibrated with the current most accurate mass values for particles and nuclei. The recalibrated $B_\Lambda$ are systematically larger (except in the case of $^6_\Lambda$He) than the original published values by about 100 keV. The effect of this level of recalibration is most significant for light hypernuclei, especially for the hypertriton. Our recalibrated $B_\Lambda$ place new constraints on theoretical studies of the strong force, of the structure of hypernuclei, and of neutron star interiors. Although this paper provides better $B_\Lambda$ estimates by recalibrating early measurements using the modern masses of particles and nuclei, the latter may also suffer from significant systematic uncertainties, such as from the energy-range relation in emulsion and emulsion density{\cite{RE_1970,Davis_2005, CSB_2017, private_communication}}. To further improve constraints on theoretical research, more precise measurements of fundamental properties of hypernuclei, like mass and binding energy, are highly desirable. More precise measurements can be expected in the near future, as a result of the on-going phase-II of the Beam Energy Scan program at RHIC, the high resolution spectroscopic experiments at Jefferson Lab{\cite{future_JLab}} in the US, and experiments at the Mainz Microton (MAMI) in Germany, while further progress will be made possible by measurements at the High Intensity Accelerator Facility (HIAF) under construction in China{\cite{HIAF_2017}}, at the Facility for Antiproton and Ion Research (FAIR) under construction in Germany, at the Japan Proton Accelerator Research Complex (J-PARC), and at the Nuclear Spectroscopic Telescope Array (NuSTAR) in the US.

\section{Acknowledgements}
We thank Prof. John Millener for insightful discussions and for sharing with us the private communication with Prof. Don Davis and the Woods-Saxon potential data.
 
\end{multicols}

\vspace{-1mm}
\centerline{\rule{80mm}{0.1pt}}
\vspace{2mm}

\begin{multicols}{2}

\end{multicols}

\clearpage

%\end{CJK*}
\end{CJK}
\end{document}

%% file: table1.tex
\begin{table}[H]
\centering
\caption{The masses of elementary particles and nuclei, as used in past publications and in 2019. The nuclear masses used in 1967 and 1968 are taken from Refs. {\cite{Nucleimass1954, Nucleimass1960_1, Nucleimass1965}}; those used in 1973 are taken from Ref. {\cite{Nucleimass1971}}, while those used in 2019 are taken from Refs. {\cite{Nucleimass_2017_1,Nucleimass_2017_2}}. All masses are in units of MeV/$c^{2}$.}
\label{table1}
\begin{tabular}{|c|c|c|c|c|}
\hline
 Particle/Nucleus & NPB1 (1967){\cite{NPB1_1967}} & NPB4 (1968){\cite{NPB4_1968}} & NPB52 (1973){\cite{NPB52_1973}} & 2019             \\ \hline
$\pi^{-}$      & 139.59{\cite{Mayeur1966}}   & 139.58{\cite{PDG_1967}}     & 139.58{\cite{PDG_1972}}       & 139.57{\cite{PDG_2018}}  \\ \hline
p              & 938.26{\cite{Mayeur1966}}   & 938.26{\cite{PDG_1967}}     & 938.26{\cite{PDG_1972}}       & 938.27{\cite{PDG_2018}}  \\ \hline
$\Lambda$      & 1115.44{\cite{NPB4_1968}}   & 1115.57{\cite{NPB4_1968}}   & 1115.57{\cite{NPB52_1973}}    & 1115.68{\cite{PDG_2018}} \\ \hline
d              & 1875.51                     & 1875.51                     & 1875.63                       & 1875.61                  \\ \hline
t              & 2808.76                     & 2808.76                     & 2808.95                       & 2808.92                  \\ \hline
$^{3}$He       & 2808.23                     & 2808.23                     & 2808.42                       & 2808.39                  \\ \hline
$^{4}$He       & 3727.17                     & 3727.17                     & 3727.42                       & 3727.38                  \\ \hline
$^{5}$He       & 4667.64                     & 4667.64                     & 4667.89                       & 4667.68                  \\ \hline
$^{6}$He       & 5605.22                     & 5605.22                     & 5605.60                       & 5605.53                  \\ \hline
$^{6}$Li       & 5601.20                     & 5601.20                     & 5601.58                       & 5601.52                  \\ \hline
$^{6}$Be       & 5604.97                     & 5604.97                     & 5605.35                       & 5605.30                  \\ \hline
$^{7}$Li       & 6533.46                     & 6533.46                     & 6533.90                       & 6533.83                  \\ \hline
$^{7}$Be       & 6533.81                     & 6533.81                     & 6534.25                       & 6534.18                  \\ \hline
$^{8}$Li       & 7470.94                     & 7470.94                     & 7471.45                       & 7471.36                  \\ \hline
$^{8}$Be       & 7454.43                     & 7454.43                     & 7454.93                       & 7454.85                  \\ \hline
$^{8}$B        & 7471.90                     & 7471.90                     & 7472.40                       & 7472.32                  \\ \hline
$^{9}$Be       & 8392.28                     & 8392.28                     & 8392.84                       & 8392.75                  \\ \hline
$^{9}$B        & 8392.83                     & 8392.83                     & 8393.40                       & 8393.31                  \\ \hline
$^{10}$Be      & 9324.97                     & 9324.97                     & 9325.60                       & 9325.50                  \\ \hline
$^{10}$B       & 9323.91                     & 9323.91                     & 9324.54                       & 9324.44                  \\ \hline
$^{11}$B       & 10251.96                    & 10251.96                    & 10252.66                      & 10252.55                 \\ \hline
$^{11}$C       & 10253.43                    & 10253.43                    & 10254.13                      & 10254.02                 \\ \hline
$^{12}$C       & 11174.23                    & 11174.23                    & 11174.98                      & 11174.86                 \\ \hline
$^{13}$C       & 12108.79                    & 12108.79                    & 12109.61                      & 12109.48                 \\ \hline
$^{13}$N       & 12110.50                    & 12110.50                    & 12111.32                      & 12111.19                 \\ \hline
$^{14}$N       & 13039.46                    & 13039.46                    & 13040.34                      & 13040.20                 \\ \hline
$^{15}$O       & 13970.39                    & 13970.39                    & 13971.33                      & 13971.18                 \\ \hline
\end{tabular}
\end{table}

%% file: table2.tex
\begin{table}[H]
\centering
\caption{The $Q_{0}$ and $\Delta Q_{0}$ values for the year indicated at the top of each column.  $\Delta Q_{0}$ denotes $Q_{0}$ in 2019 minus $Q_{0}$ in the specified year. All $Q_{0}$ and $\Delta Q_{0}$ values are in units of MeV/$c^{2}$.}

\footnotesize
{

\label{table2}
\begin{tabular}{|c| l |c|c|c|c|c|c|c|}

\hline
\multirow{2}{*}{Hypernucleus}            & \multicolumn{1}{c|}{\multirow{2}{*}{Decay modes}} & \multicolumn{2}{c|}{NPB1 (1967){\cite{NPB1_1967}}} & \multicolumn{2}{c|}{NPB4 (1968){\cite{NPB4_1968}}} & \multicolumn{2}{c|}{NPB52 (1973){\cite{NPB52_1973}}} & 2019    \\ \cline{3-9} 
                                        &\multicolumn{1}{c|}{}                              & $Q_{0}$     & $\Delta Q_{0}$                     & $Q_{0}$         & $\Delta Q_{0}$                 & $Q_{0}$          & $\Delta Q_{0}$         & $Q_{0}$      \\ \hline
\multirow{2}{*}{$^{3}_{\Lambda}$H}      & $\pi^{-}$ + $^{3}$He                              & 43.13       & 0.20                               & 43.27           & 0.06                           & 43.20            &  0.13                  & 43.33            \\
                                        & $\pi^{-}$ + p + d                                 & 37.59       & 0.25                               & 37.73           & 0.11                           & 37.73            &  0.11                  & 37.84            \\ \hline
\multirow{3}{*}{$^{4}_{\Lambda}$H}      & $\pi^{-}$ + $^{4}$He                              & 57.44       & 0.21                               & 57.58           & 0.07                           & 57.52            &  0.13                  & 57.65            \\
                                        & $\pi^{-}$ + p + t                                 & 37.59       & 0.25                               & 37.73           & 0.11                           & 37.73            &  0.11                  & 37.84            \\
                                        & $\pi^{-}$ + d + d                                 & 33.59       & 0.22                               & 33.73           & 0.08                           & 33.68            &  0.13                  & 33.81            \\ \hline
\multirow{2}{*}{$^{4}_{\Lambda}$He}     & $\pi^{-}$ + p + $^{3}$He                          & 37.59       & 0.25                               & 37.73           & 0.11                           & 37.73            &  0.11                  & 37.84            \\
                                        & $\pi^{-}$ + p + p + d                             & 32.05       & 0.30                               & 32.19           & 0.16                           & 32.26            &  0.09                  & 32.35            \\ \hline
\multirow{4}{*}{$^{5}_{\Lambda}$He}     & $\pi^{-}$ + p + $^{4}$He                          & 37.59       & 0.25                               & 37.73           & 0.11                           & 37.73            &  0.11                  & 37.84            \\
                                        & $\pi^{-}$ + d + $^{3}$He                          & 19.28       & 0.21                               & 19.42           & 0.07                           & 19.36            &  0.13                  & 19.49            \\
                                        & $\pi^{-}$ + p + p + t                             & 17.74       & 0.29                               & 17.88           & 0.15                           & 17.94            &  0.09                  & 18.03            \\
                                        & $\pi^{-}$ + p + d + d                             & 13.74       & 0.26                               & 13.88           & 0.12                           & 13.89            &  0.11                  & 14.00            \\ \hline
$^{6}_{\Lambda}$He                      & $\pi^{-}$ + d + $^{4}$He                          & 40.81       & -0.01                              & 40.95           & -0.15                          & 40.83            &  -0.03                 & 40.80             \\ \hline
\multirow{4}{*}{$^{7}_{\Lambda}$He}     & $\pi^{-}$ + $^{7}$Li                              & 47.61       & 0.20                               & 47.75           & 0.06                           & 47.69            &  0.12                  & 47.81            \\
                                        & $\pi^{-}$ + p + $^{6}$He                          & 37.59       & 0.25                               & 37.73           & 0.11                           & 37.73            &  0.11                  & 37.84            \\
                                        & $\pi^{-}$ + t + $^{4}$He                          & 45.14       & 0.20                               & 45.28           & 0.06                           & 45.22            &  0.12                  & 45.34            \\
                                        & $\pi^{-}$ + p + t + t                             & 25.29       & 0.24                               & 25.43           & 0.10                           & 25.43            &  0.10                  & 25.53            \\ \hline
\multirow{3}{*}{$^{7}_{\Lambda}$Li}     & $\pi^{-}$ + p + $^{6}$Li                          & 37.59       & 0.25                               & 37.73           & 0.11                           & 37.73            &  0.11                  & 37.84            \\
                                        & $\pi^{-}$ + $^{3}$He + $^{4}$He                   & 41.65       & 0.21                               & 41.79           & 0.07                           & 41.73            &  0.13                  & 41.86            \\
                                        & $\pi^{-}$ + p + d + $^{4}$He                      & 36.11       & 0.26                               & 36.25           & 0.12                           & 36.26            &  0.11                  & 36.37            \\ \hline
$^{7}_{\Lambda}$Be                      & $\pi^{-}$ + p + p + p + $^{4}$He                  & 38.87       & 0.35                               & 39.01           & 0.21                           & 39.14            &  0.08                  & 39.22            \\ \hline
\multirow{4}{*}{$^{8}_{\Lambda}$Li}     & $\pi^{-}$ + $^{4}$He + $^{4}$He                   & 54.97       & 0.21                               & 55.11           & 0.07                           & 55.05            &  0.13                  & 55.18            \\
                                        & $\pi^{-}$ +  p + t + $^{4}$He                     & 35.12       & 0.25                               & 35.26           & 0.11                           & 35.26            &  0.11                  & 35.37            \\
                                        & $\pi^{-}$ +  d + d + $^{4}$He                     & 31.12       & 0.22                               & 31.26           & 0.08                           & 31.21            &  0.13                  & 31.34            \\
                                        & $\pi^{-}$ + d + $^{6}$Li                          & 32.60       & 0.21                               & 32.74           & 0.07                           & 32.68            &  0.13                  & 32.81            \\ \hline
\multirow{5}{*}{$^{8}_{\Lambda}$Be}     & $\pi^{-}$ +  $^{8}$B                              & 37.76       & 0.21                               & 37.90           & 0.07                           & 37.84            &  0.13                  & 37.97            \\
                                        & $\pi^{-}$ +  p + $^{7}$Be                         & 37.59       & 0.25                               & 37.73           & 0.11                           & 37.73            &  0.11                  & 37.84            \\
                                        & $\pi^{-}$ +  p + $^{3}$He + $^{4}$He              & 36.00       & 0.25                               & 36.14           & 0.11                           & 36.14            &  0.11                  & 36.25            \\
                                        & $\pi^{-}$ + p + p + $^{6}$Li                      & 31.94       & 0.29                               & 32.08           & 0.15                           & 32.14            &  0.09                  & 32.23            \\
                                        & $\pi^{-}$ + p + p + d + $^{4}$He                  & 30.46       & 0.30                               & 30.60           & 0.16                           & 30.67            &  0.09                  & 30.76            \\ \hline
\multirow{3}{*}{$^{9}_{\Lambda}$Li}     & $\pi^{-}$ + $^{9}$Be                              & 54.51       & 0.21                               & 54.65           & 0.07                           & 54.60            &  0.12                  & 54.72            \\
                                        & $\pi^{-}$ +  p + $^{8}$Li                         & 37.59       & 0.25                               & 37.73           & 0.11                           & 37.73            &  0.11                  & 37.84            \\
                                        & $\pi^{-}$ +  t + $^{6}$Li                         & 36.83       & 0.20                               & 36.97           & 0.06                           & 36.91            &  0.12                  & 37.03            \\ \hline
\multirow{2}{*}{$^{9}_{\Lambda}$Be}     & $\pi^{-}$ + $^{9}$B                               & 37.45       & 0.20                               & 37.59           & 0.06                           & 37.52            &  0.13                  & 37.65            \\
                                        & $\pi^{-}$ +  p + $^{4}$He + $^{4}$He              & 37.68       & 0.25                               & 37.82           & 0.11                           & 37.82            &  0.11                  & 37.93            \\ \hline
\multirow{2}{*}{$^{9}_{\Lambda}$B}      & $\pi^{-}$ +  p + $^{8}$B                          & 37.59       & 0.25                               & 37.73           & 0.11                           & 37.73            &  0.11                  & 37.84            \\
                                        & $\pi^{-}$ +  p + p + p + $^{6}$Li                 & 31.77       & 0.33                               & 31.91           & 0.19                           & 32.03            &  0.07                  & 32.10            \\ \hline
$^{10}_{\Lambda}$Be                     & $\pi^{-}$ + p + p + $^{8}$Li                      & 20.67       & 0.29                               & 20.81           & 0.15                           & 20.86            &  0.10                  & 20.96            \\ \hline
\multirow{2}{*}{$^{10}_{\Lambda}$B}     & $\pi^{-}$ + p + $^{9}$B                           & 37.59       & 0.25                               & 37.73           & 0.11                           & 37.73            &  0.11                  & 37.84            \\
                                        & $\pi^{-}$ + p + p + $^{4}$He + $^{4}$He           & 37.82       & 0.30                               & 37.96           & 0.16                           & 38.03            &  0.09                  & 38.12            \\ \hline
\multirow{6}{*}{$^{11}_{\Lambda}$B}     & $\pi^{-}$ +  $^{11}$C                             & 46.33       & 0.20                               & 46.47           & 0.06                           & 46.40            &  0.13                  & 46.53            \\
                                        & $\pi^{-}$ + p + d + $^{4}$He + $^{4}$He           & 31.65       & 0.26                               & 31.79           & 0.12                           & 31.80            &  0.11                  & 31.91            \\
                                        & $\pi^{-}$ + $^{4}$He + $^{7}$Be                   & 38.78       & 0.21                               & 38.92           & 0.07                           & 38.86            &  0.13                  & 38.99            \\
                                        & $\pi^{-}$ + $^{3}$He + $^{4}$He + $^{4}$He        & 37.19       & 0.21                               & 37.33           & 0.07                           & 37.27            &  0.13                  & 37.40            \\
                                        & $\pi^{-}$ + p + $^{4}$He + $^{6}$Li               & 33.13       & 0.25                               & 33.27           & 0.11                           & 33.27            &  0.11                  & 33.38            \\
                                        & $\pi^{-}$ + t + $^{8}$B                           & 19.10       & 0.21                               & 19.24           & 0.07                           & 19.18            &  0.13                  & 19.31            \\ \hline
$^{12}_{\Lambda}$B                      & $\pi^{-}$ + $^{4}$He +  $^{4}$He + $^{4}$He       & 46.30       & 0.22                               & 46.44           & 0.08                           & 46.39            &  0.13                  & 46.52            \\ \hline
\multirow{2}{*}{$^{13}_{\Lambda}$C}     & $\pi^{-}$ + $^{13}$N                              & 39.58       & 0.20                               & 39.72           & 0.06                           & 39.65            &  0.13                  & 39.78            \\
                                        & $\pi^{-}$ + p + $^{4}$He + $^{4}$He + $^{4}$He    & 30.31       & 0.25                               & 30.45           & 0.11                           & 30.45            &  0.11                  & 30.56            \\ \hline
$^{15}_{\Lambda}$N                      & $\pi^{-}$ + $^{15}$O                              & 44.92       & 0.21                               & 45.06           & 0.07                           & 45.00            &  0.13                  & 45.13            \\ \hline
\end{tabular}

}

\end{table}

%% file: table3.tex
\begin{table}[H]
\centering
\caption{The original and recalibrated $\Lambda$ separation energy for hypernuclei in 1967{\cite{NPB1_1967}}, 1968{\cite{NPB4_1968}}, and 1973{\cite{NPB52_1973}}. The listed errors are the reported statistical uncertainties only, and the recalibrated $\Lambda$ separation energies should be considered as subject to the same errors as the original measurements. The $\Lambda$ separation energies are in units of MeV.}
\label{table3}
\begin{tabular}{|c|c|c|c|c|c|c|}
\hline
\multirow{2}{*}{Hypernucleus} & \multicolumn{2}{c|}{NPB1 (1967){\cite{NPB1_1967}}} & \multicolumn{2}{c|}{NPB4 (1968){\cite{NPB4_1968}}}& \multicolumn{2}{c|}{NPB52 (1973){\cite{NPB52_1973}}} \\ \cline{2-7}
                   & Original            & Recalibrated   & Original            & Recalibrated      & Original                & Recalibrated      \\ \hline
$^{3}_{\Lambda}$H  & 0.20 $\pm$ 0.12    & 0.41           & 0.01 $\pm$ 0.07    & 0.08              & 0.15 $\pm$ 0.08        & 0.27              \\ \hline
$^{4}_{\Lambda}$H  & 2.13 $\pm$ 0.06    & 2.35           & 2.23 $\pm$ 0.03    & 2.31              & 2.08 $\pm$ 0.06        & 2.20              \\ \hline
$^{4}_{\Lambda}$He & 2.20 $\pm$ 0.06    & 2.45           & 2.36 $\pm$ 0.04    & 2.47              & 2.42 $\pm$ 0.04        & 2.53              \\ \hline
$^{5}_{\Lambda}$He & 3.08 $\pm$ 0.03    & 3.33           & 3.08 $\pm$ 0.02    & 3.19              & 3.17 $\pm$ 0.02        & 3.28              \\ \hline
$^{6}_{\Lambda}$He & 4.09 $\pm$ 0.27    & 4.08           & 4.38 $\pm$ 0.19    & 4.23              & 4.42 $\pm$ 0.13        & 4.39              \\ \hline
$^{7}_{\Lambda}$He & 4.67 $\pm$ 0.28    & 4.88           & 4.25 $\pm$ 0.25    & 4.34              & No data                & No data           \\ \hline
$^{7}_{\Lambda}$Li & 5.46 $\pm$ 0.12    & 5.68           & 5.60 $\pm$ 0.07    & 5.67              & 5.64 $\pm$ 0.04        & 5.77              \\ \hline
$^{7}_{\Lambda}$Be & 5.36 $\pm$ 0.23    & 5.71           & 5.06 $\pm$ 0.19    & 5.27              & 5.09 $\pm$ 0.11        & 5.17              \\ \hline
$^{8}_{\Lambda}$Li & 6.72 $\pm$ 0.08    & 6.93           & 6.84 $\pm$ 0.06    & 6.91              & 6.81 $\pm$ 0.03        & 6.94              \\ \hline
$^{8}_{\Lambda}$Be & 6.67 $\pm$ 0.16    & 6.89           & 6.87 $\pm$ 0.08    & 6.95              & 6.91 $\pm$ 0.07        & 7.02              \\ \hline
$^{9}_{\Lambda}$Li & 8.27 $\pm$ 0.18    & 8.49           & 8.23 $\pm$ 0.19    & 8.34              & 8.59 $\pm$ 0.17        & 8.70              \\ \hline
$^{9}_{\Lambda}$Be & 6.66 $\pm$ 0.08    & 6.88           & 6.62 $\pm$ 0.05    & 6.68              & 6.80 $\pm$ 0.03        & 6.93              \\ \hline
$^{9}_{\Lambda}$B  & No data            & No data        & No data            & No data           & 7.89 $\pm$ 0.15        & 7.98              \\ \hline
$^{10}_{\Lambda}$Be& No data            & No data        & No data            & No data           & 9.30 $\pm$ 0.26        & 9.40              \\ \hline
$^{10}_{\Lambda}$B & No data            & No data        & No data            & No data           & 8.82 $\pm$ 0.12        & 8.93              \\ \hline
$^{11}_{\Lambda}$B & 10.30 $\pm$ 0.14   & 10.51          & 9.99 $\pm$ 0.18    & 10.11             & 10.24 $\pm$ 0.06       & 10.37             \\ \hline
$^{12}_{\Lambda}$B & 11.26 $\pm$ 0.16   & 11.48          & 10.95 $\pm$ 0.16   & 11.03             & 11.45 $\pm$ 0.07       & 11.58             \\ \hline
$^{13}_{\Lambda}$C & 10.51 $\pm$ 0.51   & 10.71          & No data            & No data           & 11.45 $\pm$ 0.12       & 11.57             \\ \hline
$^{15}_{\Lambda}$N & No data            & No data        & No data            & No data           & 13.59 $\pm$ 0.14       & 13.72             \\ \hline
\end{tabular}
\end{table}

%% file: table4.tex
\begin{table}[H]
\centering
\caption{Comparisons between $\Lambda$ separation energy ($B_{\Lambda}$) for each listed hypernucleus and the binding energy of the last neutron ($S_{n}$) and proton ($S_{p}$) in the corresponding nucleus with the same $A$ and $Z$. The $B_{\Lambda}$ values for hypernuclei with $A \leq 15$ are the recalibrated $B_{\Lambda}$ for 1973 (except in the case of $^{7}_{\Lambda}$He, where the recalibrated 1968 number is used), while data for hypernuclei with $A > 15$ are from Ref.{\cite{strangeness_2016}}. The $S_{n}$ and $S_{p}$ values are taken from a database maintained by the International Atomic Energy Agency (IAEA){\cite{IAEA}}. The $B_{\Lambda}$, $S_{n}$, and $S_{p}$ are in units of MeV.}
\label{table4}
\begin{tabular}{|c|c|c|c|c|c|c|c|}
\hline
              & $^{3}_{\Lambda}$H (t)         & $^{4}_{\Lambda}$He ($^{4}$He) & $^{5}_{\Lambda}$He ($^{5}$He) & $^{6}_{\Lambda}$He ($^{6}$He) & $^{7}_{\Lambda}$He ($^{7}$He)     & $^{7}_{\Lambda}$Li ($^{7}$Li)     & $^{7}_{\Lambda}$Be ($^{7}$Be)   \\ \hline
$B_{\Lambda}$ & 0.27 $\pm$ 0.08               & 2.53 $\pm$ 0.04               & 3.28 $\pm$ 0.02               & 4.39 $\pm$ 0.13               & 4.34 $\pm$ 0.25                   & 5.77 $\pm$ 0.04                   & 5.17 $\pm$ 0.11                 \\
$S_{n}$       & 6.26                          & 20.58                         & -0.74 $\pm$ 0.02              & 1.71 $\pm$ 0.02               & -0.41 $\pm$ 0.01                  & 7.25                              & 10.68                           \\
$S_{p}$       & No data                    & 19.81                         & 20.68 $\pm$ 0.10              & 22.59 $\pm$ 0.09              & 23.09 $\pm$ 0.25                  & 9.97                              & 5.61                            \\ \hline
              & $^{8}_{\Lambda}$Li ($^{8}$Li) & $^{8}_{\Lambda}$Be ($^{8}$Be) & $^{9}_{\Lambda}$Li ($^{9}$Li) & $^{9}_{\Lambda}$Be ($^{9}$Be) & $^{9}_{\Lambda}$B ($^{9}$B)        & $^{10}_{\Lambda}$Be ($^{10}$Be)   & $^{10}_{\Lambda}$B ($^{10}$B)   \\ \hline
$B_{\Lambda}$ & 6.94 $\pm$ 0.03               & 7.02 $\pm$ 0.07               & 8.70 $\pm$ 0.17               & 6.93 $\pm$ 0.03               & 7.98 $\pm$ 0.15                   & 9.40 $\pm$ 0.26                   & 8.93 $\pm$ 0.12                 \\
$S_{n}$       & 2.03                          & 18.90                         & 4.06                          & 1.66                          & 18.58                             & 6.81                              & 8.44                            \\
$S_{p}$       & 12.42                         & 17.25                         & 13.94                         & 16.89                         & -0.19                             & 19.64                             & 6.59                            \\ \hline
              & $^{11}_{\Lambda}$B ($^{11}$B) & $^{12}_{\Lambda}$B ($^{12}$B) & $^{13}_{\Lambda}$C ($^{13}$C) & $^{15}_{\Lambda}$N ($^{15}$N) & $^{16}_{\Lambda}$N ($^{16}$N)     & $^{16}_{\Lambda}$O ($^{16}$O)     & $^{28}_{\Lambda}$Si ($^{28}$Si) \\ \hline
$B_{\Lambda}$ & 10.37 $\pm$ 0.06              & 11.58 $\pm$ 0.07              & 11.57 $\pm$ 0.12              & 13.72 $\pm$ 0.14              & 13.76 $\pm$ 0.16                  & 13.0 $\pm$ 0.2                    & 17.2 $\pm$ 0.2                  \\
$S_{n}$       & 11.45                         & 3.37                          & 4.95                          & 10.83                         & 2.49                              & 15.66                             & 17.18                           \\
$S_{p}$       & 11.23                         & 14.10                         & 17.53                         & 10.21                         & 11.48                             & 12.13                             & 11.58                           \\ \hline
              & $^{32}_{\Lambda}$S ($^{32}$S) & $^{51}_{\Lambda}$V ($^{51}$V) & $^{52}_{\Lambda}$V ($^{52}$V) & $^{89}_{\Lambda}$Y ($^{89}$Y) & $^{139}_{\Lambda}$La ($^{139}$La) & $^{208}_{\Lambda}$Pb ($^{208}$Pb) &                                 \\ \hline
$B_{\Lambda}$ & 17.5 $\pm$ 0.5                & 21.5 $\pm$ 0.6                & 21.8 $\pm$ 0.3                & 23.6 $\pm$ 0.5                & 25.1 $\pm$ 1.2                    & 26.9 $\pm$ 0.8                    &                                 \\
$S_{n}$       & 15.04                         & 11.05                         & 7.31                          & 11.48                         & 8.78                              & 7.37                              &                                 \\
$S_{p}$       & 8.86                          & 8.06                          & 9.00                          & 7.08                          & 6.25                              & 8.00                              &                                 \\ \hline
\end{tabular}
\end{table}